\documentclass{article}
\usepackage[T1]{fontenc} 
\usepackage[utf8]{inputenc} 
\usepackage{ismir,amsmath,cite,url}
\usepackage{graphicx}
\usepackage{color}
\usepackage{booktabs}
\usepackage{amssymb,multirow}
\usepackage{etoolbox,siunitx}
\usepackage{xcolor,colortbl}
\usepackage{subcaption}

\renewcommand{\bfseries}{\fontseries{b}\selectfont}
\newrobustcmd{\B}{\bfseries}

\title{Music Separation Enhancement \\ with Generative Modeling}

\multauthor
{Noah Schaffer$^{\star 1}$ \hspace{1cm} Boaz Cogan$^{\star 1}$ \thanks{$^{\star}$ \text{Equal contribution}} \hspace{1cm} Ethan Manilow$^1$} { \bfseries{Max Morrison$^1$ \hspace{1cm} Prem Seetharaman$^2$ \hspace{1cm} Bryan Pardo$^1$}\\
 $^1$ Interactive Audio Lab, Northwestern University, Evanston IL, USA\\
$^2$ Descript, Inc. \\
}

\def\authorname{N. Schaffer, B. Cogan, E. Manilow, M. Morrison, P. Seetharaman, and B. Pardo}

\usepackage[bookmarks=false,pdfauthor={\authorname},pdfsubject={\papersubject},hidelinks]{hyperref}

\begin{document}

\maketitle
\begin{abstract}
Despite phenomenal progress in recent years, state-of-the-art music separation systems produce source estimates with significant perceptual shortcomings, such as adding extraneous noise or removing harmonics. We propose a post-processing model (the Make it Sound Good (MSG) post-processor) to enhance the output of music source separation systems. We apply our post-processing model to state-of-the-art waveform-based and spectrogram-based music source separators, including a separator unseen by MSG during training. Our analysis of the errors produced by source separators shows that waveform models tend to introduce more high-frequency noise, while spectrogram models tend to lose transients and high frequency content. We introduce objective measures to quantify both kinds of errors and show MSG improves the source reconstruction of both kinds of errors. Crowdsourced subjective evaluations demonstrate that human listeners prefer source estimates of bass and drums that have been post-processed by MSG. 


\end{abstract}
\section{Introduction}
\label{sec:introduction}

Audio source separation is the problem of isolating a sound producing source (e.g., a singer) or group of sources (e.g., a backing band) in an audio scene (e.g., a music recording). Source separation is a core problem in computer audition that can facilitate music remixing and other Music Information Retrieval (MIR) tasks such as music instrument labeling \cite{vincent2004instrument,woodruff2006remixing} and transcription \cite{plumbley2002automatic,manilow2020simultaneous}. 

Current state-of-the-art source separation systems often produce source estimates that contain perceptible artifacts, such as high-frequency noise, source leaking (e.g., drum hits heard in the bass source estimate), unnatural transients, or missing overtones. For many downstream tasks in MIR or music creation, it is preferable for source separators to minimize these errors. Given that we have observed these artifacts to be endemic to the separators themselves, we propose an additional post-processing step to clean up the initial outputs of these separators.

\begin{figure*}[t]
  \centering
  \includegraphics[width=\linewidth]{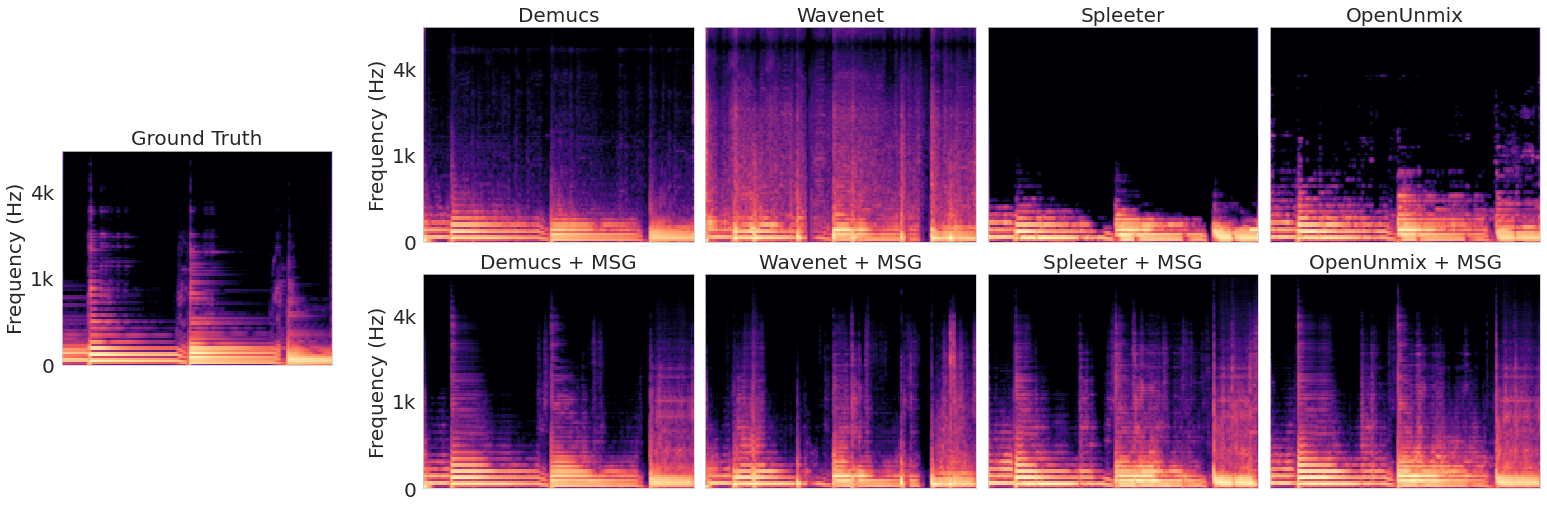}
  \caption{Spectrograms of ground-truth (left), source estimates (top), and MSG output (bottom) for the bass source. MSG is able to simultaneously infer missing frequencies and remove noise from the output of common source separation systems.}
  \label{fig:hero_image}
\end{figure*}

In this work, we introduce Make it Sound Good (MSG), a post-processing neural network for enhancing the quality of music source separation. MSG combines elements of off-the-shelf architectures from generative modeling tasks in speech vocoding and denoising to enhance the output of pre-trained source separation models in both the waveform and spectrogram domains. 

The main contributions of this work are:
\begin{itemize}
    \setlength\itemsep{-0.25em}
    \item A source separation post-processor (MSG) that performs imputation and denoising to enhance the output of both waveform and spectrogram models for music audio source separation.
    \item A subjective listener study that confirms MSG improves the perceptual quality of bass and drum source estimates on a set of five separation models, including one on which it was not trained.
    \item An in-depth exploration of the kinds of errors produced by different classes of source separators and how MSG affects these errors.
\end{itemize}


Audio examples and code can be found at
\href{https://interactiveaudiolab.github.io/project/msg.html}{\url{https://interactiveaudiolab.github.io/project/msg.html}}
.

\section{Related Work}
\label{sec:related_work}


Deep learning is the dominant approach for music source separation. For example, all entries to the 2021 Sony Music Demixing Challenge \cite{mitsufuji2021music} were deep learning based separators. Most separators fall into one of two classes. \textit{Waveform models} \cite{luo2018tasnet,defossez2019music,lluis2018end,stoller2018wave} take audio waveform input and produce an audio waveform for each separated source. \textit{Spectrogram models} \cite{reddy2004soft, weiss2006estimating, rafii2011simple,liutkus2012adaptive,hershey2016deep,luo2017deep,stoter2019open,hennequin2020spleeter,takahashi2020d3net} take a mixture spectrogram as input and output a mask to apply to the spectrogram for each source being separated. Despite the recent successes of these deep learning methods, state of the art systems continue to exhibit perceptible artifacts in their outputs. We show in Section \ref{sec:source_separation_artifacts} that waveform  models tend to introduce more high-frequency noise, while spectrogram models tend to lose transients and high frequency content.



Recent works in adversarial audio synthesis \cite{donahue2018adversarial,kumar2019melgan,binkowski2019high,engel2019gansynth,yamamoto2020parallel,kong2020hifi,jang2021univnet} and end-to-end speech enhancement \cite{pascual2017segan,fu2019metricgan,su2020hifi,pandey2020densely,su2021hifi,fu2021metricgan+,andreev2022hifi++} show that the adversarial loss of Generative Adversarial Networks (GANs) \cite{goodfellow2014generative} is effective in generating high-fidelity audio. 
Although such systems have been effective in audio synthesis and denoising, no previous work has explored using these systems for enhancing source separation output. Recent work has also shown that adversarial loss is effective for training source separation systems ~\cite{guso2022loss, stoller2018adversarial}, however this work does not look at using adversarial loss to enhance existing separation output. 

While many recent works for speech enhancement have been proposed, music enhancement (e.g., denoising and artifact removal) is less common. There are only a few recent works in music denoising \cite{li2020learning,moliner2022two,imort2022removing}
and bandwidth expansion \cite{sulun2020filter,kim2019bandwidth}. Some work has also looked at potential causes of~\cite{pons2021upsamplingartifacts} and remedies for~\cite{pons2021upsampling} audio artifacts in untrained source separation networks. Our work is most similar to Kandpal et. al \cite{kandpal2022music}, who proposed a generative model that can enhance the audio quality of a low-quality music recording taken on a consumer device. We are unaware of a prior system for enhancing the output of trained music separation systems.


\section{``Make It Sound Good'' Post-Processor}
\label{sec:MSG}

Here we describe our ``Make it Sound Good'' (MSG) post-processor, which perceptually improves a source estimate by removing artifacts that the separator introduced and imputing elements the separator omitted. We use the adversarial loss of Generative Adversarial Networks (GANs)~\cite{goodfellow2014generative} due to its success denoising many types of audio.

The generator of MSG is a waveform-to-waveform U-Net with 1D convolutions. This is very similar to the Demucs v2~\cite{defossez2019music} architecture with the exception that Demucs has two BLSTM layers at the bottleneck, which we omit.

We train the generator using three loss functions. The first is the LSGAN~\cite{mao2017least} generator loss,
\begin{equation}
    \label{eq:gen_loss}
    L_G = \frac{1}{K} \sum_{k=1}^{K} \mathbb{E} \left[ \left(D_k(G(\hat{s}))-1\right)^2\right],
\end{equation}
where $\hat{s}$ is the raw source estimate from the separator, $D_k$ is the $k$-th discriminator, K is the total number of discriminators, and $G$ is the generator. 

Next is deep feature matching loss~\cite{larsen2016autoencoding}, which is the $L_1$ distance between the intermediate activations of the discriminators on corresponding real and generated data. The last loss function we use is a multi-scale Mel-spectrogram reconstruction loss~\cite{defossez2020real, wang2019neural}, which is the average Mel reconstruction loss over three different Short-time Fourier transforms (STFTs), each of which uses different parameters for the number of STFT bins, window lengths, and hop sizes. 

We use two types of discriminators: the multi-period discriminators from HiFi-GAN ~\cite{kong2020hifi} and the multi-resolution spectrogram discriminators from UnivNet~\cite{jang2021univnet}. The multi-period discriminators operate on the waveform, and reshapes the waveform to a 2D tensor with a prime-valued stride before processing the reshaped waveform with 2D convolutional layers. The multi-resolution spectrogram discriminators process a spectrogram with different STFT window sizes (see Section \ref{subsection:training}). We use five multi-period discriminators with strides $[2,3,5,7,11]$, respectively. We also use three multi-resolution discriminators with FFT windows $[512,1024,2048]$, for a total of eight discriminators. Each discriminator uses the LSGAN~\cite{mao2017least} loss,
\begin{equation}
    \label{eq:discr_loss}
    L_D = \mathbb{E} \left[\left(D(s)-1\right)^2 + (D(G(\tilde{s})))^2\right],
\end{equation}
where $\tilde{s}$ is the cleaned up source estimate from the MSG generator and $s$ is the ground-truth source audio. Further details on the discriminator architectures are provided in the original papers ~\cite{kong2020hifi, jang2021univnet}.


We use an adversarial loss typical of Generative Adversarial Networks, but do not condition on a random input vector. This produces a deterministic model that is not technically generative. However, prior work has shown that the unique mode-selecting behaviors of these adversarial loss models are highly effective for densely-conditioned generative modeling tasks such as vocoding~\cite{kumar2019melgan,kong2020hifi,jang2021univnet} and speech denoising and enhancement~\cite{pascual2017segan,fu2019metricgan,su2020hifi,pandey2020densely,su2021hifi,fu2021metricgan+,andreev2022hifi++}. Our goal is not an exact reconstruction of ground truth, but an output that is perceptually improved. This means a distribution of viable outputs exists and the task can be framed as one of generative modeling.


\section{Experimental Validation}
\label{sec:experimental_design}

We conducted experiments to understand whether MSG perceptually enhances the raw output of a set of music source separation models. The remainder of this section is devoted to outlining the details of our experiments.


\subsection{Models}
\label{sec:separators}

We trained MSG post-processors using the output of four existing source separation models as the input to MSG. We train on two waveform-based separators: Demucs v2~\cite{defossez2019music} and Wavenet~\cite{lluis2018end}; and two spectrogram-based separators: Spleeter~\cite{hennequin2020spleeter} and OpenUnmix~\cite{stoter2019open}. To investigate whether MSG can learn to correct the artifacts of different separators, we created one enhancement model that is trained and evaluated on all four separators instead of creating separator-specific models. We evaluated the MSG post-processor using the output of each of the four separators on a held out test set (see Section~\ref{sec:data}). Furthermore, to understand whether MSG can also reduce the artifacts produced by an unseen separator, we evaluate our post-processor on the output of a fifth separator that it was never trained on: Hybrid Demucs (v3)~\cite{defossez2021hybrid}, which operates in both waveform and spectral domains. For all separation models we used the trained, frozen weights released by the authors, with no alterations. We refer readers to the papers on each separator for architectural and training details.

\subsection{Data}
\label{sec:data}

All experiments were run with the MUSDB18 dataset~\cite{musdb18}. MUSDB18 contains 150 songs: 100 in the training set and 50 in the test set. Each song in MUSDB18 has a full  mixture and isolated source audio stems for vocals, bass,  drums and a fourth catch-all category called ``other''. We omitted this catch-all category because we find that attempting to enhance many instruments at once with the same model greatly increases the difficulty of the task. We performed source separation on every song in MUSDB18 using all five of our source separators, producing source estimates of bass, drums and vocals. 

The input audio was peak normalized before passing it through the network. Since Wavenet operates at 16 kHz we use this sample rate. We downsampled all systems to 16 kHz so that there was a uniform sample rate across all separation models. Here, we focus solely on enhancement and leave the task of bandwidth extension for future work. Thus, the output of MSG on all systems was at a sample rate of 16 kHz.

\subsection{Training}
\label{subsection:training}
We trained one MSG model on the MUSDB18 training set for each source class (bass, drums, or vocals). Each model was trained using source estimates from four separators (Demucs v2, Wavenet, Spleeter, and OpenUnmix) as input and the ground-truth sources as training targets. We segmented the audio into 1-second clips and rejected silent clips where the ground-truth source has an RMS below -60dB FS, resulting in over 100,000 training examples per model. On each training iteration, we randomly swapped the input data with ground-truth with a $10\%$ probability. This encourages the model to leave high-quality audio unaltered. 

We computed the three resolutions of STFTs passed to our multi-resolution spectrogram discriminators (Section~\ref{sec:MSG}) using window sizes of 512, 1024, and 2048 samples and hop sizes of 128, 256, and 512 samples, respectively. We used one Adam optimizer\cite{kingma2014adam} for the generator and another for the discriminator. We used a learning rate of 2e-4 and beta values of $(.5, .9)$.
To find suitable loss weights for all 3 types of losses on the generator (LSGAN loss, deep matching loss, and multi-scale spectral loss; see Section~\ref{sec:MSG}), we solved a least squares equation to weigh all loss terms equally for the first 1k training iterations. After that, we froze those weights applied to the losses for the remainder of training.



\subsection{Subjective evaluation}
\label{sec:subjective}


\begin{figure}[ht]
  \centering

  \includegraphics[width=\linewidth]{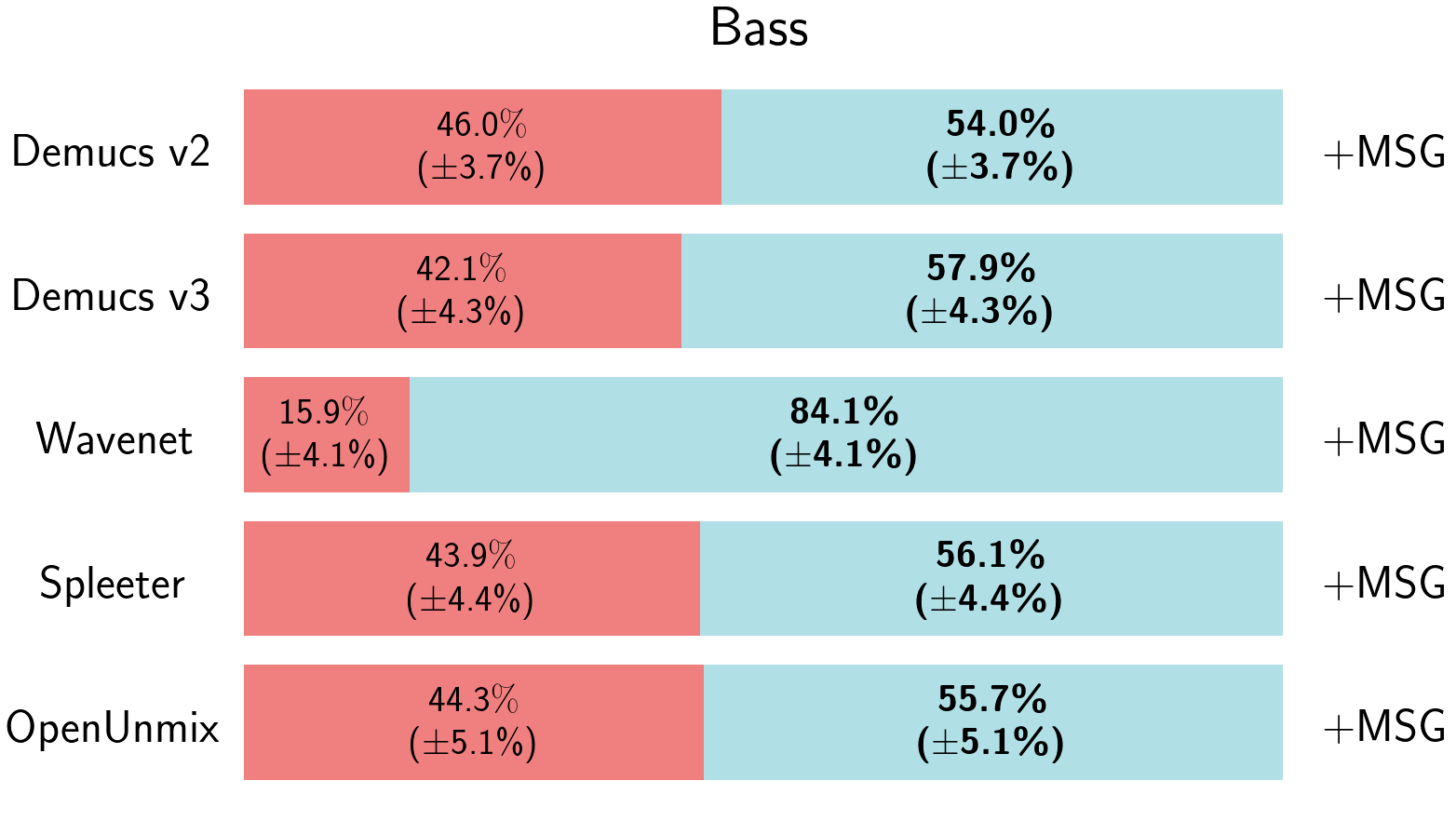}
  \includegraphics[width=\linewidth]{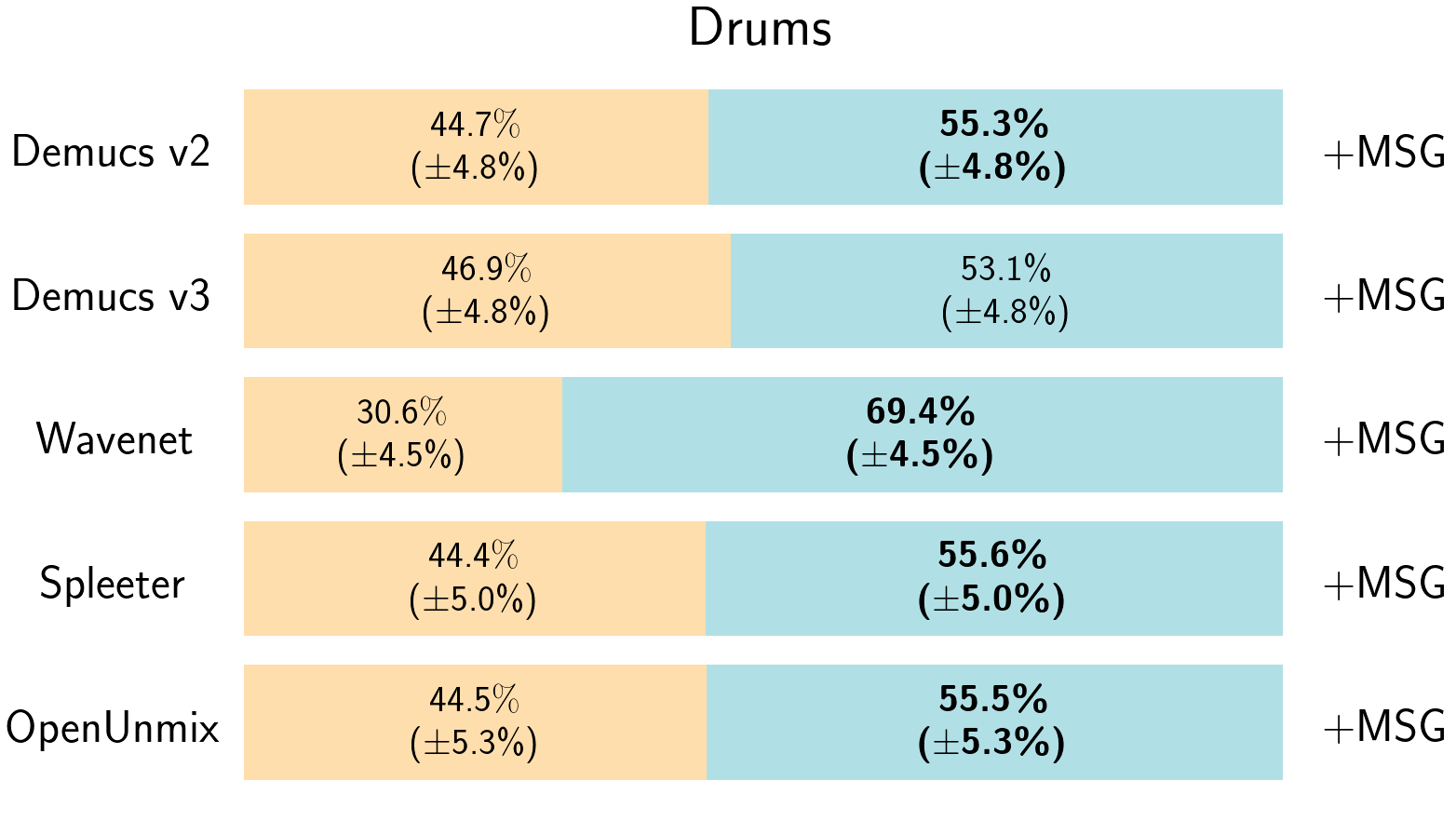}
  \includegraphics[width=\linewidth]{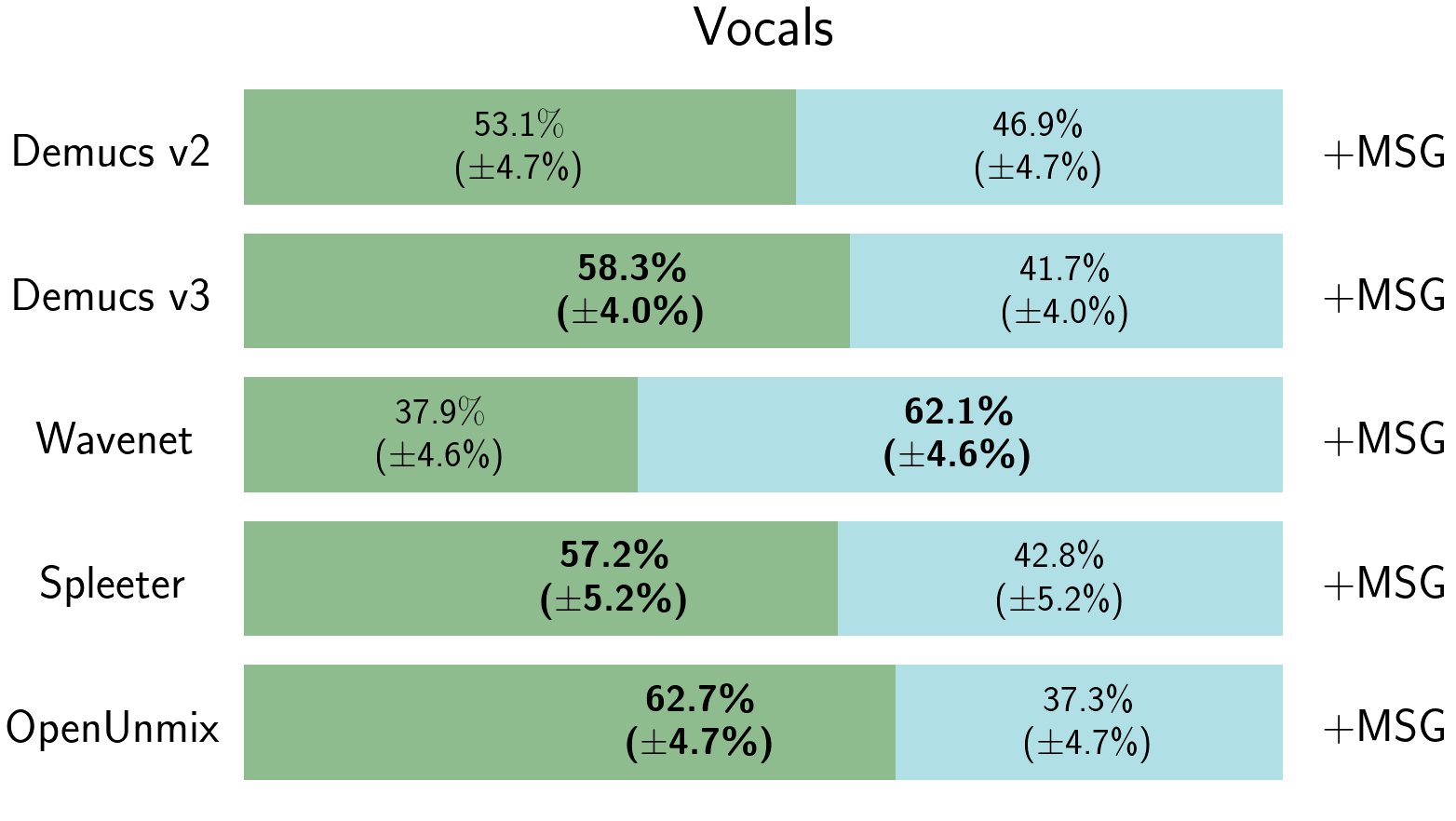}
  \caption{Subjective pairwise test results for bass, drums, and vocals. Each row contains the percent of listeners selecting that option as higher quality in a two-way forced choice listening test. A bold-faced value indicates a statistically significant difference.}
  \label{fig:pairwise}
\end{figure}

The goal of our research is to improve the perceptual quality of source separation output.  Therefore, we evaluate our MSG post-processor using a crowdsourced subjective evaluation (Section~\ref{sec:subjective}) rather than reporting an objective metric like Signal-to-Distortion Ratio (SDR) \cite{vincent2006performance, le2019sdr}, since widely-used objective metrics for source separation are imperfect proxies for human perception \cite{fox2007modeling, cartwright2016fast,cartwright2018crowdsourced, gupta2015perceptual,cano2016evaluation,le2019sdr,guso2022loss}.

For evaluation data, we used one seven-second segment from each of the 50 songs from the MUSDB18 test set. We performed source separation on each seven-second segment using each of the five source separation systems (Section~\ref{sec:separators}) to create source estimates of bass, drums, and vocals. Each output was then processed with MSG, resulting in 50 matched pairs for each combination of separator and source class: the raw output, and the output processed by MSG. 

There are 15 unique combinations of the five separators and three sources (bass, drums, and vocals). For each combination, we performed a two-way forced-choice listening test between the raw output and the output processed by MSG. We initially recruited 20 participants for each test and omitted responses from participants that failed a prescreening listening test. This resulted in a minimum number of 15 participants in any test. 
Each participant evaluated 25 randomly-selected pairs from the 50 examples for that combination of source and separator. 

A two-tailed binomial test was performed where the null hypothesis was that there was no difference between MSG-enhanced and raw separator output.  
If the results of a particular test showed no difference (i.e., $p < 0.05$) we recruited an additional 10 participants to see if a difference could be determined. 

For each pairwise comparison, participants were given the following instructions, where <\textit{source}> is one of ``bass'', ``drums'' or ``vocals'':

\begin{quote}
Listen to both recordings of a <\textit{source}>. After listening to both, select the recording that sounds like a higher-quality \textit{<source>}. The higher-quality recording is the one that is more natural sounding, or has fewer audio artifacts (e.g., noise, clicks, or other instruments).
\end{quote}

We used Reproducible Subjective Evaluation (ReSEval)\cite{morrison2022reproducible} to set up our listener studies. We recruited participants via Amazon Mechanical Turk (MTurk). Our participants were US residents at least 18 years old that completed 20 or more tasks on MTurk with an approval rating of at least 97\%. Participants who passed the listening test and completed our evaluation were paid $\$3.00$.

For each of the 15 tests, we collected between 308 and 696 pairwise evaluations from between 15 and 30 participants who passed the prescreening listening test. The number of evaluations is not a multiple of 25 because a few participants did not finish all 25 examples in their set of pairwise evaluations.

\subsubsection{Subjective evaluation results}

Each listening test evaluates one combination of source class and source separator. Figure~\ref{fig:pairwise} shows the results for each of the 15 tests. Listeners preferred the MSG output to the raw source separator output in 11 out of 15 combinations of separator and source. This difference was statistically significant (using a binomial test) in 10 of the 11 combinations. Listeners preferred the quality of separators with MSG on bass source estimates and had a moderate preference for MSG on drums. For vocals, listeners had a slight preference for the source estimate without MSG. 

MSG performed best on the Wavenet separator, where it significantly improved the perceptual audio quality of all sources. MSG was also able to improve on the quality of source estimates of a separator not seen during training, Demucs v3, for bass sources---as well as an improvement on Demucs v3 drum sources that was not statistically significant. Note that Demucs v3 is a hybrid approach that operates in both the waveform and spectrogram domains. Our performance on vocals indicates that MSG is not able to enhance the quality of the source estimate of vocals.    
We are unaware of prior work that attempts to enhance source separation output, let alone a separated vocal source.
Vocal separation has been optimized by many years of existing research, potentially leaving less room for a post-processing system to improve compared to, e.g., drum and bass sources.

\begin{figure}[t]
  \centering
  \includegraphics[width=\linewidth]{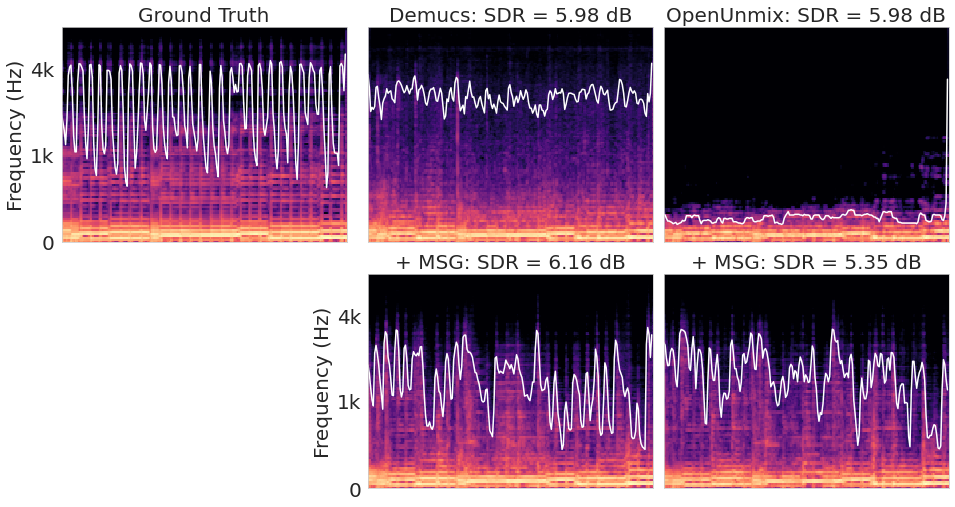}
  \caption{Spectrograms of the ground truth (left) and source estimates from Demucs v2 and OpenUnmix (top) and corresponding MSG output (bottom) for the bass source. The $98\%$ spectral rolloff frequency is overlaid in white.}
  \label{fig:sdr-rolloff}
\end{figure}

\section{Further Analysis}
\label{sec:source_separation_artifacts}

\begin{figure*}[t]
  \centering
  \includegraphics[width=\linewidth]{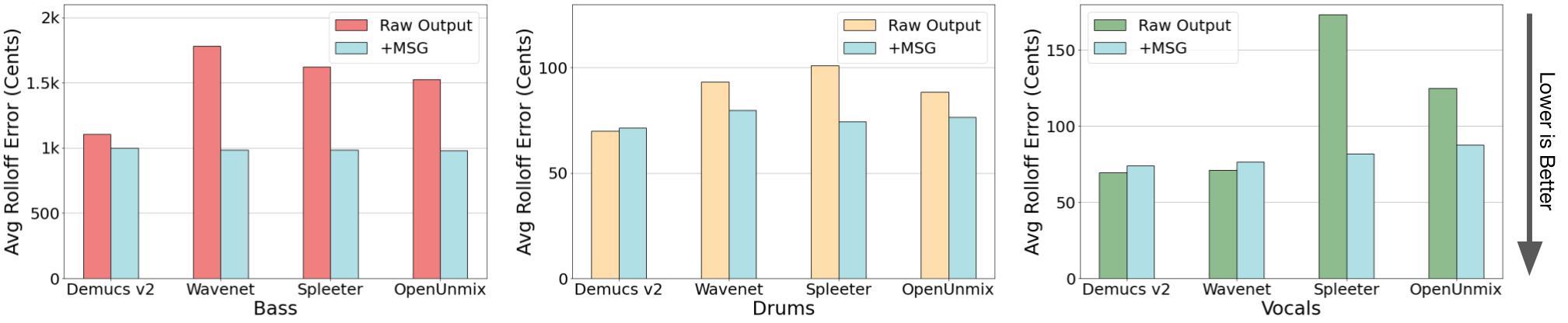}
  \caption{Mean spectral rolloff error for bass, drums, and vocals for separators with and without MSG post-processing.}
  \label{fig:rolloff-error}
\end{figure*}

Our listener study indicates whether a separation is relatively ``good'' or ``bad'', but it does not clarify  \textit{why}  one separated source is better or worse than another. Similarly, the widely-used Signal-to-Distortion Ratio (SDR) ~\cite{vincent2006performance, le2019sdr}, as well as the related SIR and SAR, are not designed to capture the specific types of errors we focus on in this work. See, for example, the top row of Figure \ref{fig:sdr-rolloff}, which shows source estimates produced by Demucs v2 and OpenUnmix for the same bass source from the same mixture. Demucs adds additional high-frequency noise not present in the ground truth, while OpenUnmix removes many of the upper harmonics. Visually, the difference between these two systems is plain, however their SDR values (using \texttt{mus\_eval}~\cite{SiSEC18}) are equal to two decimal places: \textit{5.98 dB!} 

In this section, we examine the output of the four state-of-the-art separation systems used in the training of MSG models: (Demucs v2 \cite{defossez2019music}, Wavenet \cite{lluis2018end}, Spleeter \cite{hennequin2020spleeter}, and OpenUnmix \cite{stoter2019open}), as well as the MSG-processed outputs for those four systems. As before, we use the MUSDB18~\cite{musdb18} test set, and we omit the ``other'' source.

Anecdotally, we have noted that waveform separators tend to add extra high-frequency noise and spectrogram separators tend to remove high-frequency partials, especially in bass estimates (see Figure~\ref{fig:hero_image}). Spectrogram separators also tend to smooth out transients. While these are not the only issues that current separation systems exhibit, the rest of this section will be dedicated to analysis of these two issues. 

\subsection{Added and Missing Frequency Content}

Following our anecdotal observation that waveform separators tend to add extra high-frequency noise and spectrogram separators tend to remove high-frequency partials, we seek to formalize these notions. 

One statistic that can be a good proxy for whether a source estimate has excess high-frequency content or is missing desirable high-frequency content is spectral rolloff. For a given time frame in a spectrogram, the spectral rolloff at $X\%$ is the frequency below which $X\%$ of the energy of the signal lies. For example, the white line on each spectrogram in Figure \ref{fig:sdr-rolloff} shows the spectral rolloff at $98\%$. 

For every song in the MUSDB18~\cite{musdb18} test set, we compute the spectral rolloff at $98\%$ every 32 ms (a hop size of 512 samples at 16 kHz) for every ground truth isolated source, every estimate produced by one of the four training separators and every MSG-enhanced source estimate. To calculate our statistics, we omit any frames that have an RMS less than $-40$ dBFS, in the ground-truth source, so as not to examine rolloff in relatively silent regions. We report the error between a source estimate's rolloff and a ground-truth source's rolloff in cents, which is $1200 \times (\log_2x - \log_2y)$, where $x$ and $y$ are rolloff frequencies in Hz. We chose to use cents over Hz because it better correlates to how humans perceive audio. 

In Figure~\ref{fig:rolloff-error}, we show the mean error, in cents,  between the ground truth rolloff and the source estimate's rolloff. We see that the source estimates of vocals and drums have spectral rolloff errors on the order 100-200 cents, whereas source estimates of bass have errors of roughly 1000 cents. MSG reduces this error for all separators on bass sources, for three of the four separators on drums, and two of the four separators on vocals.

\begin{figure}[t]
  \centering
  \includegraphics[width=\linewidth]{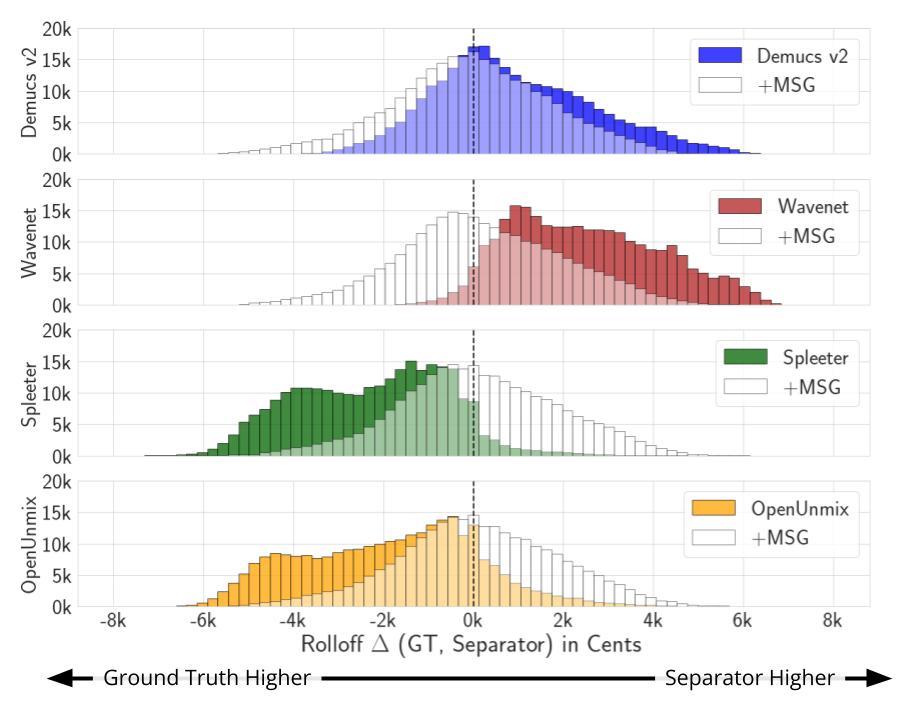}
  \caption{Histogram of the difference in spectral rolloff values between a given separator's bass estimate and ground truth bass source over the MUSDB18 test set. The vertical dotted line shows the desired difference of 0. MSG reduces the difference between the rolloff values of source estimate and ground-truth.}
  \label{fig:bass_rolloff_histogram}
\end{figure}

Because the bass estimates have such large errors, we examine them further in Figure~\ref{fig:bass_rolloff_histogram}, where we show a histogram of the per-frame differences between the spectral rolloff of source estimates and ground truth. The top two rows of the histogram show results for the two waveform separators (i.e., Demucs v2 and Wavenet), which each show an error distribution that is strongly skewed towards positive error. This corroborates our observation of high-frequency noise introduced by waveform separators, as shown in the bass spectrogram in Figure~\ref{fig:sdr-rolloff}.  The bottom two rows show the error distribution for two spectrogram separators, Spleeter and OpenUnmix. Both exhibit an error distribution for spectral rolloff that is strongly skewed toward negative values. This quantifies the effect illustrated in Figures~\ref{fig:sdr-rolloff} and~\ref{fig:hero_image}, where the spectrogram separators remove the higher partials of the bass source. 

We further observe that, when MSG is applied to the output of all four separators, the resulting error distribution is less biased and, as was already shown in Figure \ref{fig:rolloff-error}, reduces the mean error magnitude. Figure \ref{fig:hero_image} illustrates the effect of MSG on a single bass example, showing improved spectral rolloff reconstruction for both waveform and spectrogram models. 

\subsection{Improving Transient Reconstruction}

While listening to the source estimates from spectrogram separators, we noticed that the transients of source estimates for drums and bass did not sound as clear as in the ground truth source estimates. To quantify these observations, we measure the location and strength of onsets in the estimated sources relative to the ground-truth. We use librosa's \cite{mcfee2015librosa} \texttt{onset\_strength()} function \cite{bock2013maximum}, which computes the spectral flux onset strength envelope at every frame in a spectrogram. We approximate an onset by identifying every frame with a strength above a certain threshold. We select an onset strength threshold via manual tuning. We set the threshold value to a constant value of 0.75 for both bass and drums on the MUSDB18 dataset. We manually tuned this threshold to find a value that best corresponds with our perception of relevant peaks in the signal. We chose to threshold \texttt{onset\_strength()} instead of using librosa's \texttt{onset\_detect()} because we found that matching up onsets between two signals using the latter method was hard to correctly tune.

We run this onset strength thresholding on both the ground-truth source and a source estimate and then calculate the F1 between the binary threshold arrays of the raw source estimate and the MSG post-processed estimate as a proxy for how well a separator preserves transients. A true positive (TP) is when a detected onset exists at the same spectrogram frame in the ground-truth and source estimate, a false positive (FP) is when an onset is detected at a frame in the source estimate but not the ground truth, and a false negative is when an onset is detected (FN) at a frame in the ground truth but not in the source estimate. We report the F1 score of onset reconstruction, $TP / (TP + \frac{1}{2}(FP + FN))$.


\begin{table}[]
\begin{center}
\sisetup{table-format=2.2,round-mode=places,round-precision=2,table-number-alignment = center,detect-weight=true,detect-inline-weight=math}
\begin{tabular}{@{}llcSS@{}}
\toprule
 \multicolumn{2}{c}{\multirow{2}{*}{Model}} & \multirow{2}{*}{Type} & \multicolumn{2}{c}{Onset Strength F1}      \\
 & & & {Raw} & {+ MSG} \\
\midrule
 & Demucs v2 & Wave & .52 & \B .54 \\
\rowcolor{gray!12} \cellcolor{white} & Wavenet & Wave & .36 & \B .44 \\

 & Spleeter & Spec & .36 & \B .52 \\
\rowcolor{gray!12}
\multirow{-4}{*}{\cellcolor{white} Bass} & OpenUnmix & Spec & .39 & \B .49 \\
\midrule
 & Demucs v2 & Wave & \B .84 & .82 \\
\rowcolor{gray!12} \cellcolor{white} & Wavenet & Wave & .73 & \B .74 \\
\cellcolor{white} & Spleeter & Spec & .78 & \B .82 \\
\rowcolor{gray!12}
\multirow{-4}{*}{\cellcolor{white} Drums} & OpenUnmix & Spec & .78 & \B .79 \\
\midrule
 & Demucs v2 & Wave & \B .58 & .57 \\
\rowcolor{gray!12} \cellcolor{white} & Wavenet & Wave & \B .51 &  .49 \\
\cellcolor{white} & Spleeter & Spec & \B .71 &  .66 \\
\rowcolor{gray!12}
\multirow{-4}{*}{\cellcolor{white} Vocals} & OpenUnmix & Spec & .41 & \B .57 \\

\bottomrule
\end{tabular}
\end{center}
\caption{F1 scores for thresholded onset strength for bass, drums, and vocals for four separators with and without MSG post-processing. ``Raw'' means that F1 is computed between the separator's raw output and ground truth. ``+MSG'' means that MSG post-processing is applied to the raw source estimates. According to this measure, MSG is able to better preserve onsets in 7 out of 8 cases between the bass and drums sources, which most clearly demonstrate artifacts with transients.}
\label{tab:precision-recall}
\end{table}

We report the F1 scores for onset detection on bass, drums, and vocals in Table \ref{tab:precision-recall}. The results for vocals are not in favor of MSG for 3 of the 4 separators. We include the vocals results for completeness. Evaluating the transients for vocals might be slightly unusual, but the observed results align with the listener studies. In contrast with vocals, bass and drums both see improved F1 scores across multiple separators: MSG improves the F1 score in 7 out of the 8 combinations of source and separator, with the sole exception of drums separated by Demucs v2. This indicates that the ability to represent transients is generally improved by applying MSG-based post processing on bass and drums.

\section{Conclusion}
\label{sec:conclusion}

State-of-the-art music source separators create audible perceptual degredations, such as missing frequencies and transients. In this work, we propose Make it Sound Good (MSG), a post-processing neural network that leverages generative modeling to enhance the perceptual quality of music source separators. In listening studies, users prefer bass and drum source estimates produced with MSG post-processing---even on a state-of-the-art separator not seen during training. We analyze the errors of waveform-based and spectrogram-based separators with and without MSG. Without MSG, we show that waveform-based separators induce high-frequency noise and spectrogram-based separators fail to reconstruct high-frequencies in the bass source, and have trouble reconstructing transients. We measure these artifacts via spectral rolloff and onset detection and show that, for both bass and drums, MSG generally improves reconstruction of spectral rolloff and onsets of the source estimate relative to the ground-truth sources. Fruitful directions for future work include using more modern techniques for sound enhancement (e.g., diffusion models~\cite{kandpal2022music}), making post-processors for the vocals and ``other'' sources, and deeper analyses of the issues with separation systems.


\clearpage

\bibliography{ISMIRtemplate}

\end{document}